\documentclass{article}

\usepackage{arxiv}

\usepackage[utf8]{inputenc} 
\usepackage[T1]{fontenc}    
\usepackage{hyperref}       
\usepackage{url}            
\usepackage{booktabs}       
\usepackage{amsfonts}       
\usepackage{nicefrac}       
\usepackage{microtype}      
\usepackage{lipsum}
\usepackage{graphicx}
\usepackage{subfigure}
\usepackage{amsmath}
\usepackage{caption}
\title{Non-degenerate wavelength computational ghost imaging with thermal light}
\author{
  Deyang Duan\thanks{Corresponding author: duandy2015@qfnu.edu.cn}, Zhongxiao Man, Yunjie Xia\\
  College of Physics and Engineering, Qufu Normal University, Qufu 273165, China\\
  Shandong Provincial Key Laboratory of Laser Polarization and Information\\
  Technology, Research Institute of Laser, Qufu Normal University, Qufu 273165, China\\
  }

\begin{document}
\maketitle

\begin{abstract}
Non-degenerate wavelength computational ghost imaging with thermal light
source is studied theoretically and experimentally. The acquired computational
ghost images are of high quality when the wavelength of
computed light is different from the light detected by bucket detector.
Compared to the necessary light of short wavelength in previous ghost imaging,
the use of longer wavelength light is demonstrated to bring about ghost images
with higher spatial resolution, in strong atmospheric turbulence.
\end{abstract}

\section{Introduction}

Ghost imaging is a transverse active-imaging technique that exploits the
correlation between two light beams to image an object without spatially
resolving measurements of the light beam that has undergone object interaction
[1]. Over the past two decades, GI has been found to have some unique
advantages over conventional optical imaging technologies, such as
super-resolution [2,3], good imaging quality in harsh optical environment
[4,5]. Recently, the experiments of GI with X-ray and electronics sources
[6-8] have been reported, indicating that GI has become a powerful
comprehensive tool in exploring and analyzing the internal characteristics of complex
material, e.g. biomolecular structures and nanomaterials.

GI is considered to have broad applied prospects [9-13], especially in remote
sensing and laser radar. One of the key reasons that many applications have not yet
realized in practice is the complexity of the optical path structure of conventional GI.
Fortunately, in 2008, Shapiro proposed a novel GI scheme-computational ghost imaging
(CGI)[14], in which the image is reconstructed by correlating a calculated pattern
with the signal of a bucket detector located behind the object illuminated by
modulated light by spatial light modulator (SLM).
Deterministic modulation of a laser beam with an SLM can provide the signal
field used for target interrogation, while the on-target intensity pattern
needed for the reference field can then be calculated via diffraction theory [15,16].

CGI is currently the most promising imaging solution for remote sensing and laser
radar because it has only one optical path, which is different from the conventional
GI essentially. Although GI can bring about good imaging quality in harsh optical
environment, it is not negligible that the ghost image will be significantly degraded
with strong turbulence and large propagation distance. 2011, Meyers \emph{et al}
proposed turbulence-free GI, which played a crucial role in applications [4].
Previous works have shown that the resolution of non-degenerate wavelength GI is higher than that of
degenerate wavelength GI [17,18]. In addition, the source with shorter wavelength illuminating
the object is effective for ghost images with higher resolution in atmospheric
turbulence [19,20].

Non-degenerate wavelength CGI with thermal light
source relies on a modification of the conventional CGI set-up (Fig.1).
A binocular charge-coupled device (CCD) camera is employed to replace the
bucket detector, compared to the conventional CGI scheme. The binocular
CCD camera can not only output the bucket signal but also be used as a
tool to measure the distance between the SLM and the object [21,22].

\section{Theory}

In the following, we  theoretically illustrate the concept of nondegenerate
wavelength CGI with thermal light source. The collected light by the
detector is named as signal light, while the calculated patterns by the SLM is
named as reference light. To explore the properties of nondegenerate
wavelength CGI, a continuous laser beam $E_{s}$ is projected to an SLM and
the modulated light illuminates the object after propagating a distance of $z$ in free space.
The light field detected by the binocular CCD camera can be expressed as [5,23]
\begin{equation}
E_{s}(x_{s})=\frac{-1}{\lambda_{1}\sqrt{z_{1}z_{2}}}\int dudyE_{s}^{^{\prime}%
}\left(  \lambda_{s},u\right)  e^{\frac{i\pi}{\lambda_{s}z_{1}}\left(
y-u\right)  }\,T(y)e^{\frac{i\pi}{\lambda_{s}z_{2}}\left(  x_{s}-y\right)  },
\end{equation}
where $T(y)$ represents the information of object. The quantities $u$, $y$,
$x$ represent the transverse coordinates of the SLM, the object plane and the bucket detector plane,
respectively. $z_{1}$ is the distance from the SLM to the
object. $z_{2}$ represents the distance from object to the binocular CCD camera.

\begin{figure}[ht!]
\centering\includegraphics [width=7cm]{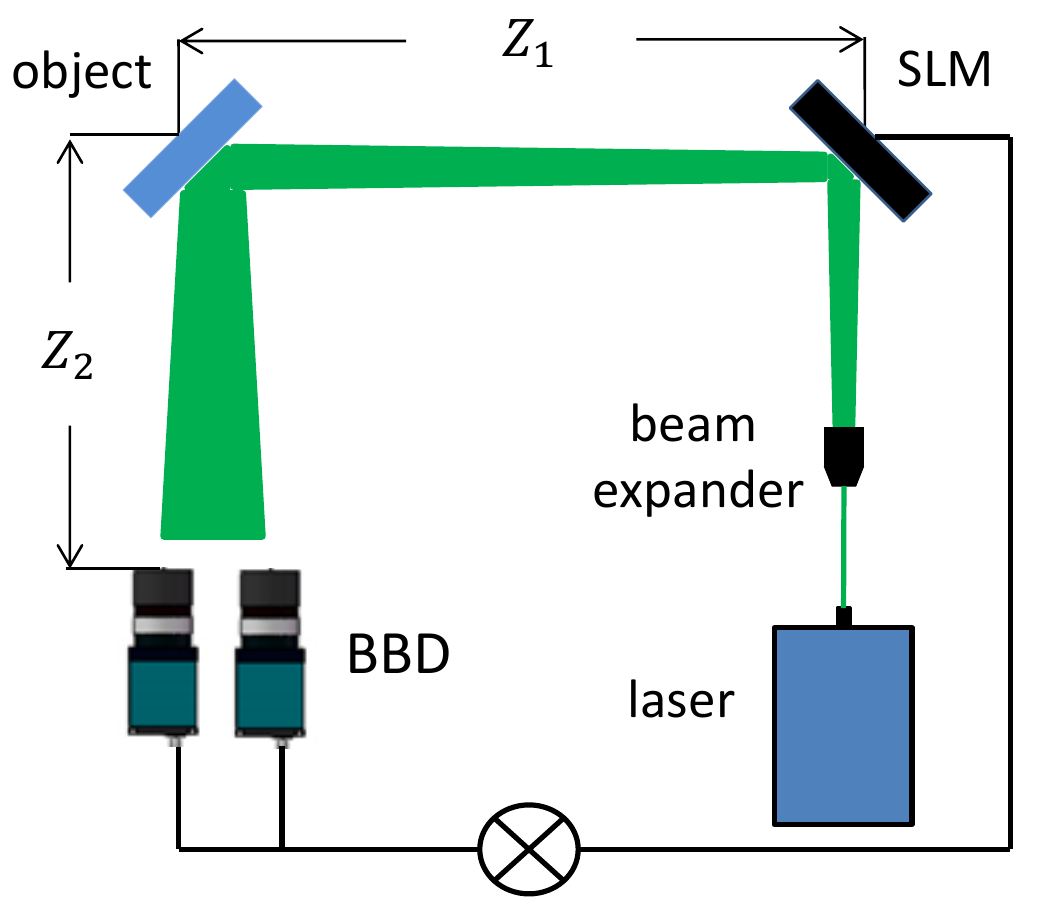}
\caption{Setup of the non-degenerate-wavelength computational ghost imaging with thermal
light source. SLM: spatial light modulator, BBD: binocular bucket detector.}
\end{figure}
In practice, $z_{1}$ and $z_{2}$ need to be measured in real time to get the calculated
light field according to the diffraction principle.
Fortunately, the distance from the object to detector can be accurately measured
by the binocular parallax principle without adding other equipments [21,22].
For simplicity and practicality, we assume $z_{1}=z_{2}=L_{1}$. In order to
obtain a nondegenerate-wavelength computational ghost image, the wavelength of
the calculated light $E_{2}$ should be different from that of the detected
light $E_{s}$, i.e., $\lambda_{s}\neq\lambda_{r}$. Furthermore, the
propagation distance $L_{2}=z_{3}$ of the calculated light is determined by
the optimal imaging conditions of the non-degenerate GI ($\lambda_{s}%
L_{1}=\lambda_{r}L_{2}$) [17]. Thus, the distribution of the computed light is
given by
\begin{equation}
E_{r}(x_{r})=\left(  \frac{-i}{\lambda_{r}\sqrt{z_{3}}}\right)  ^{1/2}\int
duE_{r}^{^{\prime}}\left(  \lambda_{r},u\right)  e^{\frac{i\pi}{\lambda
_{r}z_{3}}\left(  u-x_{r}\right)  }\,.
\end{equation}

In order to reconstruct the image, the calculated patterns
at the object plane are cross-correlated with the signal
output by the detector, i.e.,
\begin{align}
&  G(x_{s},x_{r})=\left\langle I_{s}\left(  x_{s}\right)  I_{r}\left(
x_{r}\right)  \right\rangle -\left\langle I_{s}\left(  x_{s}\right)
\right\rangle \left\langle I_{r}\left(  x_{r}\right)  \right\rangle
\nonumber\\
&  =\left\langle E_{s}^{\ast}\left(  x_{s}\right)  E_{s}\left(  x_{s}\right)
E_{r}^{\ast}\left(  x_{r}\right)  E_{r}\left(  x_{r}\right)  \right\rangle
-\left\langle E_{s}^{\ast}\left(  x_{s}\right)  E_{s}\left(  x_{s}\right)
\right\rangle \left\langle E_{r}^{\ast}\left(  x_{r}\right)  E_{r}\left(
x_{r}\right)  \right\rangle \nonumber\\
&  =\frac{1}{\lambda_{s}^{2}\lambda_{r}z_{1}z_{2}z_{3}}\int du_{1}%
du_{1}^{^{\prime}}du_{2}du_{2}^{^{\prime}}dydy^{^{\prime}}C\left(  \lambda
_{s},\lambda_{r};u_{1},u_{1}^{^{\prime}},u_{2},u_{2}^{^{\prime}}\right)
T(y)T^{\ast}(y)\\
&  \times e^{\frac{i\pi}{\lambda_{s}z_{1}}\left[  \left(  y-u_{1}\right)
^{2}-\left(  y^{^{\prime}}-u_{1}^{^{\prime}}\right)  ^{2}\right]  }%
e^{\frac{i\pi}{\lambda_{s}z_{2}}\left[  \left(  x_{s}-y\right)  ^{2}-\left(
x_{s}^{^{\prime}}-y^{^{\prime\prime}}\right)  ^{2}\right]  }e^{\frac{i\pi
}{\lambda_{r}z_{3}}\left[  \left(  u_{2}-x_{r}\right)  ^{2}-\left(
u_{2}^{^{\prime}}-x_{r}^{^{\prime}}\right)  ^{2}\right]  },\nonumber
\end{align}
where
\begin{align}
&  C\left(  \lambda_{s},\lambda_{r};u_{1},u_{1}^{^{\prime}},u_{2}%
,u_{2}^{^{\prime}}\right)  \nonumber\\
&  =\left\langle E_{s}^{^{\prime}\ast}\left(  \lambda_{s},u_{1}\right)
E_{s}^{^{\prime}}\left(  \lambda_{s},u_{1}^{^{\prime}}\right)  E_{r}%
^{^{\prime}\ast}\left(  \lambda_{r},u_{2}\right)  E_{r}^{^{\prime}}\left(
\lambda_{r},u_{2}^{^{\prime}}\right)  \right\rangle \\
&  =\left\langle E_{s}^{^{\prime}\ast}\left(  \lambda_{s}\right)
E_{s}^{^{\prime}}\left(  \lambda_{s}\right)  \right\rangle \left\langle
E_{r}^{^{\prime}\ast}\left(  \lambda_{r}\right)  E_{r}^{^{\prime}}\left(
\lambda_{r}\right)  \right\rangle \left\langle V\left(  u_{1}\right)  V^{\ast
}\left(  u_{2}^{^{\prime}}\right)  \right\rangle \left\langle V\left(
u_{2}\right)  V^{\ast}\left(  u_{1}^{^{\prime}}\right)  \right\rangle
\nonumber
\end{align}
is the intensity cross-correlation function of the light beams in the spatial
and temporal frequency domain evaluated at the output plane of the SLM.
$E_{a}\left(  \lambda_{a},u_{a}\right)  =E_{a}^{^{\prime}}\left(  \lambda
_{a}\right)  V\left(  u_{a}\right)  $, $a=s,r$. $E_{s}^{^{\prime}}$ and
$E_{r}^{^{\prime}}$ are the independent light fields at the input plane of the
SLM. In order to simplify the calculation, we assume that $\left\langle
E^{\ast}\left(  \lambda\right)  E\left(  \lambda\right)  \right\rangle
=I_{0}=1$. In addition, the SLM mask function $V\left(  u\right)  $
possesses spatial correlations that follow Gaussian statistics.%

\begin{equation}
\left\langle V\left(  u_{1}\right)  V^{\ast}\left(  u_{2}^{^{\prime}}\right)
\right\rangle =e^{-\frac{u_{1}^{2}+u_{1}^{^{\prime}2}+u_{2}^{2}+u_{2}%
^{^{\prime}2}}{4\omega}}e^{-\frac{\left(  u_{1}-u_{2}^{^{\prime}}\right)
^{2}+\left(  u_{2}-u_{1}^{^{\prime}}\right)  ^{2}}{2l_{c}^{2}}},
\end{equation}
where $\omega$ is the transverse size of the laser beam and $l_{c}$ is the
correlation parameter of the random amplitude caused by the SLM. Substituting
Eq.(5) into Eq.(3), we can thus write the ghost
image of Eq.(3) as
\begin{equation}
G(x_{r})=\frac{\pi^{2}}{\lambda_{s}\lambda_{r}z_{1}z_{3}\sqrt{ABCD}}\int
dyO\left(  y\right)  \exp\left(  \frac{S^{2}}{4A}+\frac{P^{2}}{4B}+\frac
{Q^{2}}{4C}+\frac{R^{2}}{4D}\right)  ,
\end{equation}
where,
\begin{align*}
A &  =\frac{1}{4\omega^{2}}+\frac{1}{2l_{c}^{2}}-\frac{i\pi}{\lambda_{s}z_{1}%
},B=\frac{1}{4\omega^{2}}+\frac{1}{2l_{c}^{2}}+\frac{i\pi}{\lambda_{s}z_{1}%
},\\
C &  =\frac{1}{4\omega^{2}}+\frac{1}{2l_{c}^{2}}-\frac{i\pi}{\lambda_{r}z_{3}%
}-\frac{1}{4Bl_{c}^{4}},D=\frac{1}{4\omega^{2}}+\frac{1}{2l_{c}^{2}}%
+\frac{i\pi}{\lambda_{r}z_{3}}-\frac{1}{4Al_{c}^{4}},\\
S &  =\frac{2i\pi y}{\lambda_{s}z_{1}},P=\frac{2i\pi y}{\lambda_{s}z_{1}%
},Q=\frac{i\pi y}{Bl_{c}^{2}\lambda_{s}z_{1}}-\frac{2i\pi x_{2}}{\lambda
_{r}z_{3}},R=\frac{2i\pi x_{2}}{\lambda_{r}z_{3}}-\frac{i\pi y}{Al_{c}%
^{2}\lambda_{s}z_{1}},
\end{align*}
$\left\langle T(y)T^{\ast}(y)\right\rangle =\lambda_{s}O\left(  y\right)
\delta\left(  y-y^{^{\prime}}\right)  .$ From Eq.(6), we can see that the
correlation function $G(x_{r})$ has the same form as that of two-color GI
[5,17]. Therefore, the non-degenerate wavelength thermal CGI feasible in
practice even when the difference of the wavelengths is very large.
Different from the two-color GI, as the reference light field is completely
calculated, the wavelength can be set arbitrarily,

We calculate the exponential function $\exp\left(  \frac{S^{2}}{4A}%
+\frac{P^{2}}{4B}+\frac{Q^{2}}{4C}+\frac{R^{2}}{4D}\right)  $ on the right of
Eq.(6), and then give the standard form of GI. This calculation process is
simple but cumbersome, so the result is directly given by
\begin{equation}
G(x_{r})=\frac{\pi^{2}}{\lambda_{s}\lambda_{r}z_{1}z_{3}\sqrt{ABCD}}%
\exp\left(  -\frac{x_{s}^{2}}{2W_{fov}^{2}}\right)  \int dyO\left(  y\right)
\exp\left(  -\frac{\left(  y-mx_{s}^{2}\right)  }{2W_{psf}^{2}}\right)  ,
\end{equation}
where $W_{fov}$ and $m$ represent the view of field and magnification factor
[5], respectively.
\begin{equation}
W_{psf}=Re\sqrt{\frac{1}{2\left(  K_{1}^{2}+K_{2}^{2}+K_{3}^{2}+K_{4}%
^{2}\right)  }}%
\end{equation}
is the width of point spread function (PSF).
\begin{align*}
K_{1} &  =\frac{\pi}{\sqrt{A}\lambda_{s}z_{1}},K_{2}=\frac{\pi}{\sqrt
{B}\lambda_{s}z_{1}},K_{3}=\frac{\pi}{2B\sqrt{C}\lambda_{s}z_{1}l_{c}^{2}},\\
K_{4} &  =\frac{\pi}{2\sqrt{D}\lambda_{s}z_{1}}\left(  \frac{1}{Bl_{c}^{2}%
}-\frac{1}{Al_{c}^{2}}\right)  .
\end{align*}

\begin{figure}[ht!]
\centering\includegraphics[width=10cm]{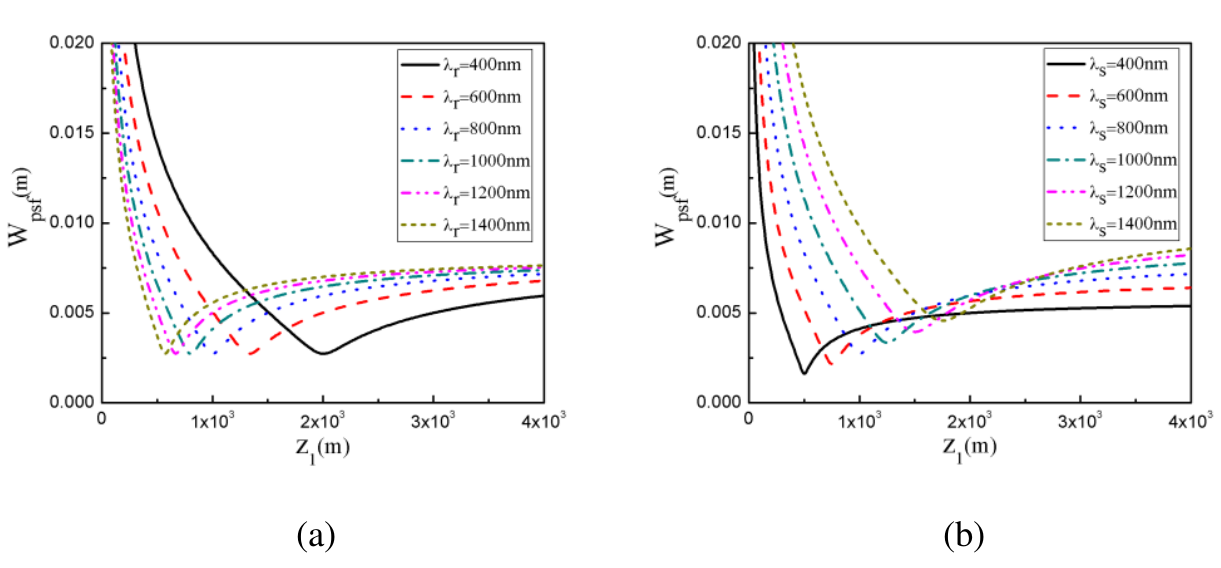}
\caption{(a) Width of the point-spread function as a function of $\lambda
_{s}=800nm$, $\lambda_{r}=400,600,800,1000,1200,1400nm$. Parameters used are
$\omega=5cm$, $l_{c}=1mm$, $z_{1}=1km$. (b)The corresponding plot of PSF as a
function of $\lambda_{s}=800nm$, $\lambda_{r}=400,600,800,1000,1200,1400nm$.}
\end{figure}

Equation (8) is plotted in Fig.2 as a function of variable $\lambda_{r}$,
$\lambda_{s}$ and $L_{1}=z_{1}$. It is seen that the width of the PSF attains
its smallest value when $\lambda_{s}L_{1}=\lambda_{r}L_{2}$, in
previous work[15]. Fig.2(a) compares the resolution of conventional CGI with
single wavelength light source (blue dot line) and nondegenerate wavelength
CGI (other kinds of lines), from which we found that there is no significant
difference between them (minimum value of PSF).
Compared to this one, the spatial resolution changes significantly as the
wavelength of signal light changes. This result has illustrated the results
that the spatial resolution of non-degenerate wavelength CGI depends more strongly on signal light
$\lambda_{s}$ than reference light $\lambda_{r}$ (Fig.2(b)).

In the case of atmospheric turbulence, based on the extended
Huygens-Fresnel integral, the propagation function from the source plane to
the detector plane is given by [5,23]
\begin{align}
E_{s}(x_{s}) &  =\frac{-1}{\lambda_{s}\sqrt{z_{1}z_{2}}}\int dudyE_{s}%
^{^{\prime}}\left(  \lambda_{s},u_{1}\right)  e^{\frac{i\pi}{\lambda_{s}z_{1}%
}\left(  y-u_{a}\right)  }\,\nonumber\\
&  \times\exp\left[  \phi_{1}(y,u)\right]  T(y)\exp\left[  \phi_{2}%
(x,y)\right]  T(y)e^{\frac{i\pi}{\lambda_{s}z_{2}}\left(  x_{s}-y\right)  },
\end{align}
where $\phi_{1}$ and $\phi_{2}$ characterize the atmospheric turbulence effects in
the SLM-to-object path and the object-to-bucket-detector path, respectively.
Because the reference arm can be computed by the simulation program according
to the diffraction theory, the computed light field is not affected by
atmospheric turbulence. Thus, the imaging function of nondegenerate-wavelength
CGI with atmospheric turbulence can be expressed as
\begin{align}
G(x_{r}) &  =\frac{\pi^{2}}{\lambda_{s}\lambda_{r}z_{1}z_{3}\sqrt{A^{^{\prime
}}B^{^{\prime}}C^{^{\prime}}D^{^{\prime}}}}\exp\left(  -\frac{x_{s}^{2}%
}{2W_{fov}^{2}}\right)  \nonumber\\
&  \times\int dyO\left(  y\right)  \exp\left(  -\frac{\left(  y-mx_{s}%
^{2}\right)  }{2W_{psf}^{2}}\right)  .
\end{align}
The PSF has the same form as Eq. 8, i.e.,
\begin{equation}
W_{psf}=Re\sqrt{\frac{1}{2\left(  K_{1}^{2}+K_{2}^{2}+K_{3}^{2}+K_{4}%
^{2}\right)  }},
\end{equation}
with
\begin{align}
A^{^{\prime}} &  =\frac{1}{4\omega^{2}}+\frac{1}{2l_{c}^{2}}+\frac{1}%
{2\rho^{2}}-\frac{i\pi}{\lambda_{s}z_{1}},B^{^{\prime}}=\frac{1}{4\omega^{2}%
}+\frac{1}{2l_{c}^{2}}+\frac{1}{2\rho^{2}}+\frac{i\pi}{\lambda_{s}z_{1}}%
-\frac{1}{4A^{^{\prime}}\rho^{4}},\nonumber\\
C^{^{\prime}} &  =\frac{1}{4\omega^{2}}+\frac{1}{2l_{c}^{2}}+\frac{1}%
{2\rho^{2}}-\frac{i\pi}{\lambda_{r}z_{3}}-\frac{1}{4B^{^{\prime}}l_{c}^{4}}\\
D^{^{\prime}} &  =\frac{1}{4\omega^{2}}+\frac{1}{2l_{c}^{2}}+\frac{i\pi
}{\lambda_{r}z_{3}}-\frac{1}{4A^{^{\prime}}l_{c}^{2}}-\frac{1}{16A^{^{\prime
}2}B^{^{\prime}}l_{c}^{4}\rho^{4}}-\frac{1}{64A^{^{\prime}2}B^{^{\prime}%
2}C^{^{\prime}}l_{c}^{8}\rho^{4}},\nonumber
\end{align}

\begin{align*}
K_{1}  &  =\frac{\pi}{\sqrt{A^{^{\prime}}}\lambda_{s}z_{1}},K_{2}=\frac{\pi
}{\sqrt{B^{^{\prime}}}\lambda_{s}z_{1}}\left(  1-\frac{1}{2A^{^{\prime}}%
\rho^{2}}\right)  ,\\
K_{3}  &  =\frac{\pi}{2B^{^{\prime}}\sqrt{C^{^{\prime}}}\lambda_{s}z_{1}%
l_{c}^{2}}\left(  1-\frac{1}{2A^{^{\prime}}\rho^{2}}\right)  ,\\
K_{4}  &  =\frac{\pi}{2\sqrt{D^{^{\prime}}}\lambda_{s}z_{1}}\left[  -\frac
{1}{A^{^{\prime}}l_{c}^{2}}+\frac{1}{B^{^{\prime}}l_{c}^{2}}\left(  1-\frac
{1}{2A^{^{\prime}}\rho^{2}}\right)  \right. \\
&  \left.  \times\left(  \frac{1}{2A^{^{\prime}}\rho^{2}}+\frac{1}%
{8A^{^{\prime}}B^{^{\prime}}C^{^{\prime}}l_{c}^{4}\rho^{2}}\right)  \right]  ,
\end{align*}
where $\rho=\left(  0.55C_{n,s}^{2}k_{s}^{2}z_{0}\right)  ^{-\frac{3}{5}}$,
$C_{n,s}^{2}$ is the turbulence strength.

\begin{figure}[ht!]
\centering\includegraphics[width=9.5cm]{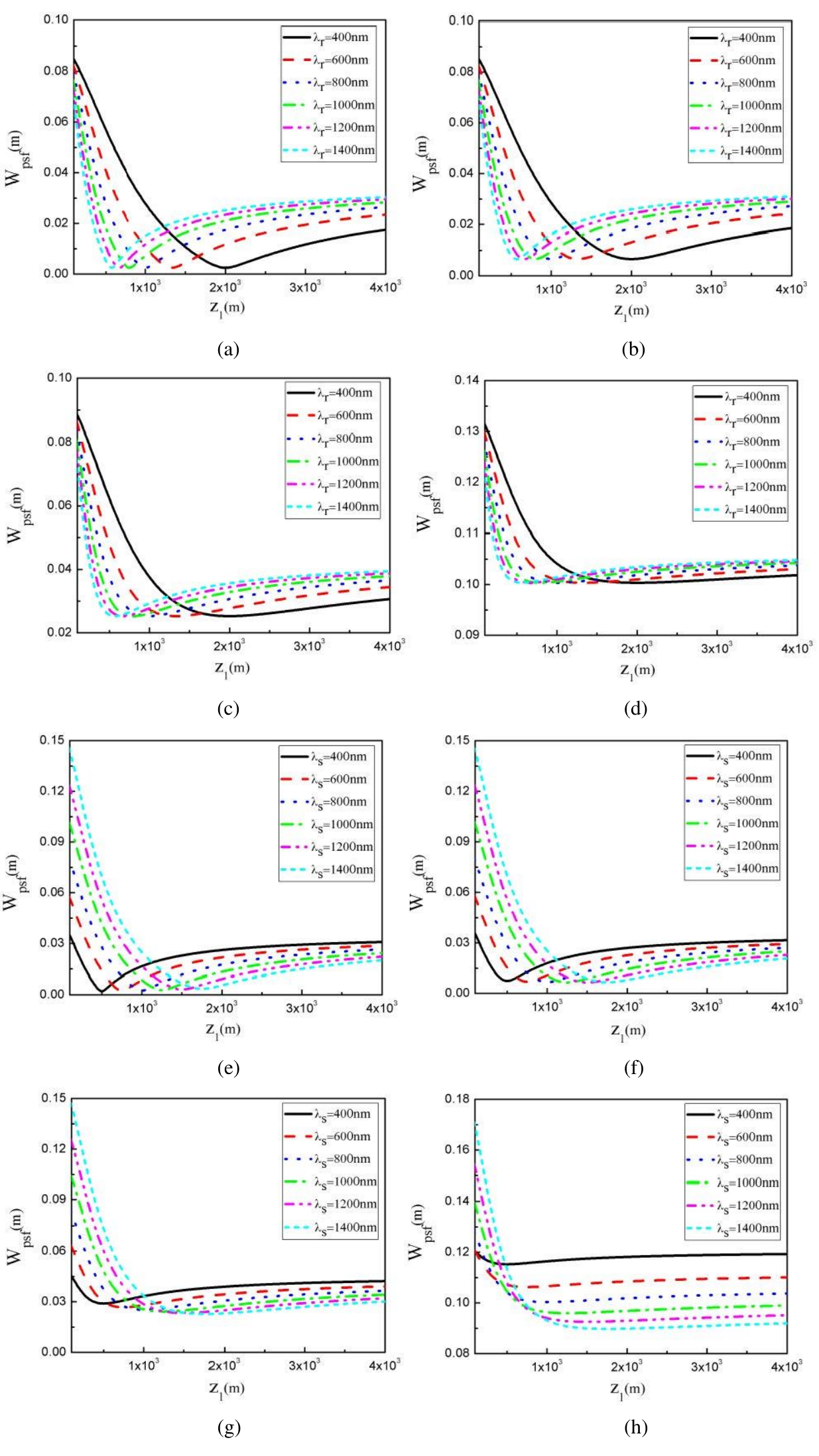}
\caption{The PSF of the nondegenerate wavelength CGI with thermal light source.
Parameters used are $\omega=5cm$, $l_{c}=1mm$, $z_{1}=1km$. (a-d) are the
curve lines for $C_{n,s}^{2}=10^{-15}$, $10^{-14}$, $10^{-13}$, $10^{-12}$,
$\lambda_{s}=800nm$, $\lambda_{r}=400,600,800,1000,1200,1400nm$. (e-h) are the
curve lines for $C_{n,s}^{2}=10^{-15}$, $10^{-14}$, $10^{-13}$, $10^{-12}$,
$\lambda_{r}=800nm$, $\lambda_{s}=400,600,800,1000,1200,1400nm$.}
\end{figure}

Equation (11) is plotted in Fig.3 as a function of $\lambda_{r}$ and
$\lambda_{s}$. We test the resolution of nondegenerate wavelength
thermal CGI with the wavelength $\lambda_{s}$ fixed and the wavelength
$\lambda_{r}$ changed. The results are shown in Fig.3(a)-3(d). The results show
that the spatial resolution of non-degenerate wavelength CGI with thermal light
is equal to that of conventional CGI in tuebulence. The spatial resolution
will be reduced with the increase of turbulence. We fix the $\lambda_{r}$ and
change $\lambda_{s}$, the results are shown in Fig.3(e)-3(h). In weak turbulence
($C_{n,s}^{2}=1\times10^{-15}$), ghost images with higher resolution can be
obtained when the light of signal arm carries shorter wavelength, which is the
same as the previous works [5,16,19]. However, in strong turbulence
($C_{n,s}^{2}=1\times10^{-12}$), ghost images with higher resolution can be
obtained when the light of signal arm carries longer wavelength, which is
different from previous works [5,16,19].
\newpage
\section{Experiments}
To experimentally demonstrate the non-degenerate wavelength CGI,
we used the imaging
setup described in the previous section to image a Rubik's Cube. The setup is
based on a two-dimensional amplitude-only ferroelectric liquid crystal spatial
light modulator (FLC-SLM, Meadowlark Optics A512-450-850), with $512\times512$
addressable $15\mu m\times15\mu m$ pixels. The light sources are CW solid-state
lasers with wavelength 532nm and 635 nm respectively. The bucket detector is replaced by
a binocular camera (CCD) with $640\times480$ addressable $3.0\mu m\times3.0\mu
m$. A $5\times5\times5cm^{3}$ object (Rubik's Cube) located at a distance
$L_{1}=100cm$ from the SLM, and a lens collects the reflected light onto the
detector. In each realization a grayscale with $512\times512$ is sent
to the SLM. The distance between the two detectors of the binocular camera is 12
cm. The measurement error of binocular ranging is about 1\%. The accuracy of
the measurement can be higher by optimizing the algorithm and calibration.

\begin{figure}[ht!]
\centering\includegraphics[width=10cm]{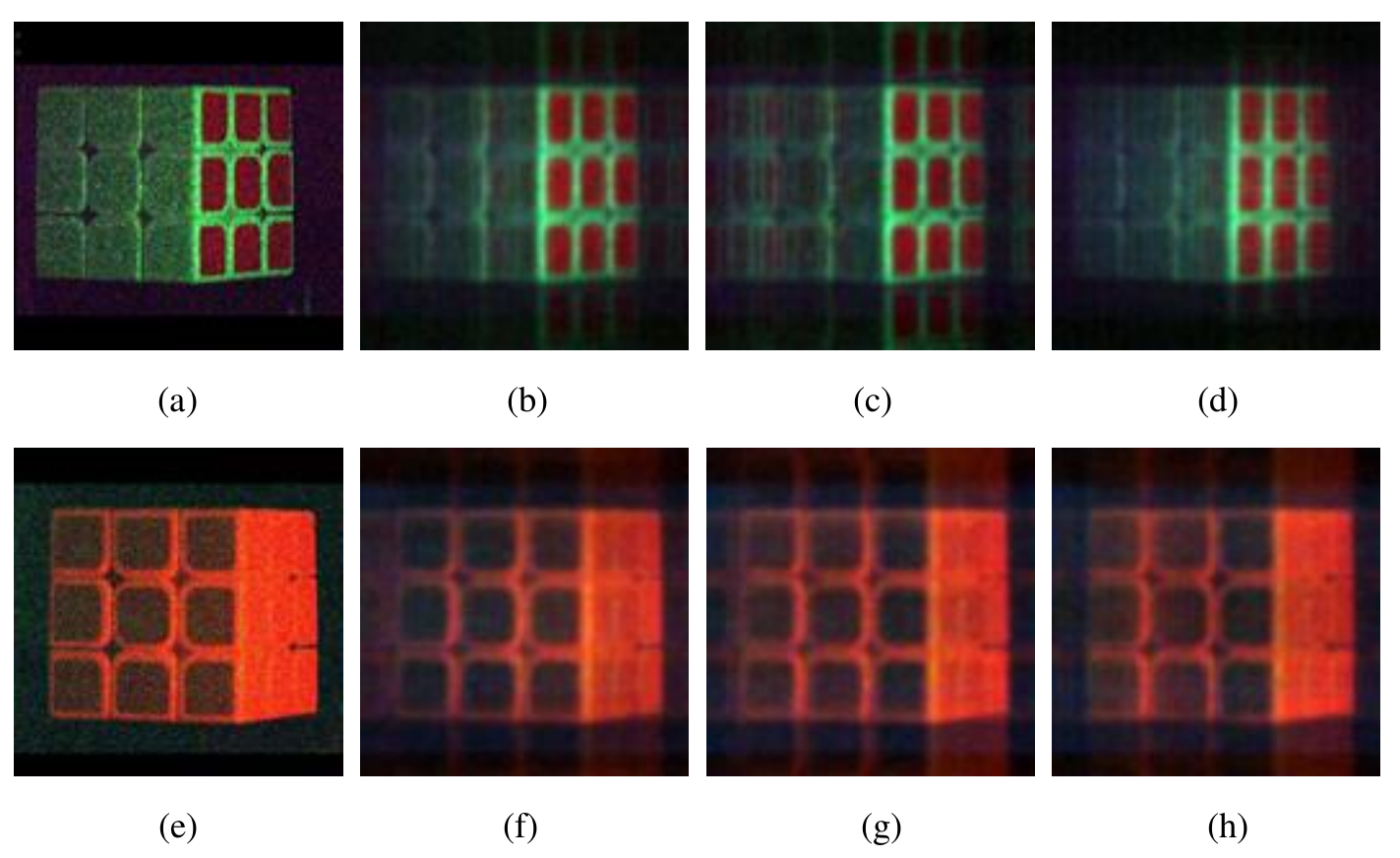}
\caption{The nondegenerate-wavelength computational ghost image with 500000
realizations. (a,e) object,(b-d) Parameters used are $\lambda_{s}=532nm$,
$\lambda_{r}=400nm,532nm,635nm$. (f-h)The corresponding results with $\lambda
_{s}=635nm$, $\lambda_{r}=400nm,532nm,635nm$.}
\end{figure}

Figure 4 compares the nondegenerate-wavelength CGI with different wavelengths
of signal light and reference light. The experiment results show that a
high-quality ghost image can be obtained even when the wavelengths of light
used in the signal and reference arms are very different. Furthermore,
changing the wavelength of the reference light does not significantly improve
the resolution of ghost image. The color of non-degenerate wavelength ghost
image depends on the light detected by the bucket detector[24,25]. It is
noteworthy that the red color in Fig.4(a)-4(d) is produced by laser-activated
coatings, which does not affect the experimental results.

\begin{figure}[ht!]
\centering\includegraphics[width=13cm]{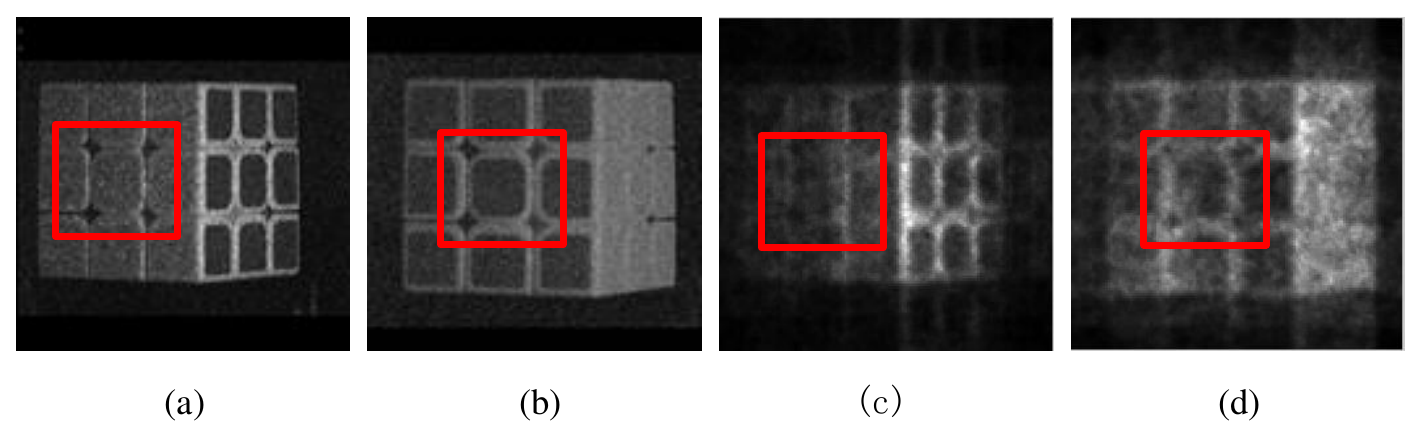}
\caption{The nondegenerate-wavelength computational ghost image with 500000
realizations. (a,b) object, (c) Parameters used are $\lambda_{r}=\lambda
_{s}=532nm$. (d)The corresponding results with $\lambda_{r}=\lambda_{s}%
=635nm$.}
\end{figure}

To verify the effect of strong atmospheric turbulence on non-degenerate wavelength CGI, turbulence is
introduced by adding heating elements at ${550}^{\circ}{\rm C}$%
underneath optical path [4]. Heating of the air causes temporal and spatial
fluctuations on its index of refraction that makes the classical image of the
object jitter about randomly on the image plane causing a \textquotedblleft
blurred\textquotedblright\ picture. Fig.5 shows that the spatial resolution
of ghost image obtained by 635 nm (Fig.5(d)) is better than that of 532 nm
(Fig.5(c)). The reason for this phenomenon is that strong turbulence has less
effect on long wavelength laser than that on short wavelength laser [26].
Further studies have shown that this conclusion applies to all types of ghost imaging.

\section{Discussions and conclusion}

Shapiro creatively proposed the concept of CGI based on the
modulation of light field by SLM. Compared with
conventional GI, the CGI is more suitable for laser radar, remote sensing and
other fields. Chan \emph{et al}. proposed two-color GI based on the modulation
effect of spatial light modulator on different wavelength light, and proved
that the resolution of ghost image depends on each of these wavelengths in
non-degenerate wavelength GI with thermal light [17]. Actually, two-color GI is
still a conventional GI rather than CGI. Here, our results show that the
resolution of non-degenerate wavelength CGI with thermal light depends
primarily on the wavelength used to illuminate the object, but not on the
calculated light. It is not difficult to understand that the correlation function of
non-degenerate wavelength CGI is the same as that of two-color GI, because
they are all based on the modulation of light field by spatial light
modulator. Thus, the spatial intensity pattern of the light on the signal arm
and computed (reference) arm is the same, which are similar to that of
conventional GI [27].

In conclusion, we have shown that nondegenerate wavelength CGI with thermal
light source is not only feasible but can also give a ghost image with high
quality. The key to nondegenerate wavelength CGI is to use the binocular
ranging principle to measure the distance of object in real time to obtain the
calculated light field. Moreover, there is no fundamental limit in the extent for
the wavelength in the reference arm as the corresponding field is completely
calculated. It is out of the restriction of the spatial light
modulator, which is better than the conventional two-color GI. More important,
the results show that the longer wavelength laser is sent to illuminate the object
in strong atmospheric turbulence, the higher resolution of ghost image can be obtained.
In addition, this scheme may have potential imaging applications for moving objects,
which will be discussed in detail in our future works.
\section*{Funding}

National Natural Science Foundation of China (11704221, 11574178, 61675115); Natural Science
Foundation of Shandong Province (ZR2016AP09, ZR2016JL005); Taishan Scholar Foundation
of Shandong Province (tsqn201812059).
\section*{Acknowledgments}
The authors wish to thank Dongfeng Shi for his contribution to resolution
calculation.

\end{document}